\documentclass[runningheads]{llncs}

\input{psfig.sty}
\usepackage{subfigure}
\usepackage{graphicx}
\usepackage{color}

\begin{document}

\pagestyle{headings}

\mainmatter

\title{Growing Trees in Internet News Groups and Forums}

\author{B. Kujawski\inst{1}$^,$\inst{2}, J. Ho{\l}yst\inst{2} \and G. J. Rodgers\inst{1}}
\authorrunning{Bernard Kujawski et al.}

\institute{Department of Mathematical Sciences, Brunel University, Uxbridge;\\
Middlesex UB8 3PH; UK \and Faculty of Physics, Warsaw University
of Technology, Koszykowa 77, 00-622 Warsaw; Poland} \maketitle

\begin{abstract}

We present an empirical study of the networks created by users
within internet news groups and forums and show that they organise
themselves into scale-free trees. The structure of these trees
depends on the topic under discussion; specialist topics have
trees with a short shallow structure whereas more universal topics
are discussed widely and have a deeper tree structure. For news
groups we find that the distribution of the time intervals between
when a message is posted and when it receives a response exhibits
a composite power-law behaviour. From our statistics we can see if
the news group or forum is free or is overseen by a moderator. The
correlation function of activity, the number of messages posted in
a given time, shows long range correlations connected with the
users' daily routines. The distribution of distances between each
message and its root is exponential for most news groups and
power-law for the forums. For both formats we find that the
relation between the supremacy ( the total number of nodes that
are \emph{under} the node $i$, including node $i$) and the degree
is linear $s(k)\sim k$, in contrast to the analytical relation for
Barab\'{a}si-Albert network.
\end{abstract}

\section{Introduction}

One of the most important features of the internet is the
opportunity it offers people to exchange opinions with one
another. Now anyone can participate in a discussion or debate
on-line and the global reach of the internet allows a single
person's opinion to be shared with people from all over the world.
Thus each of us can now be a source of information, not only for
our relatives and friends, but for the whole world. We can offer
our opinion to a very wide range of people and receive feedback on
this opinion. Thus internet discussions are potentially important
in helping to shape people's opinions and behaviour and in the
spreading of ideas and information. In this way the internet is a
medium which is very different to traditional media such as
newspapers, radio and television. The use of the internet has lead
to an explosion of interest within other academic disciplines in
phenomena such as social contagion, viral marketing and stealth
marketing. Despite this importance, scientific research into
internet discussions has been rather limited.

There have only been a few scientific papers examining internet
discussion networks. Makowiec and Bykowska \cite{makowiec}
considered the three most popular blog web pages in Poland. They
provided an analysis of the network structure of blogs and gave a
sociological explanation to the results. In related work, Zhongbao
and Changshui \cite{Zhongbao} examined the network properties of
bulletin board systems (BBS), which are similar to the news groups
examined in this paper. They \cite{Zhongbao} studied a network in
which edges were between users, and were able to identify distinct
communities within the network of users. BBS and users networks
were also studied by Goh \textit{et al.} \cite{Goh}, who found
intercommunities and intracommunities with different topological
properties. The intercommunity was a homogenous system, in which
members all knew each other, while intercommunities were
characterised by a power law degree distribution. Capocci
\textit{et al.} \cite{Capocci} investigated the largest internet
encyclopedia, Wikipedia. A bow-tie-like, scale-free structure with
almost neutral mixing was found. Only small and medium nodes
exhibited linear preferential attachment. Valverde and Sol\'{e}
\cite{Valverde} focused on technology development communities,
such as open source communities, by looking at e-mail exchanges.
Non-local growth rules based on a betweenness centrality model
were examined and compared with the empirical data. The temporal
properties of e-mail exchange groups were studied by Barab\'{a}si
\cite{Barabasi}.

Internet forums and news groups are similar to BBS networks, but
in contrast to previous work
\cite{makowiec,Zhongbao,Goh,Valverde}, here we place an edge
between messages and focus on the network of ideas or opinions
posted by users, rather than networks between the users
themselves. In this way we obtain tree like networks with a
central topic, the root node, and the surrounding threads.

In the last few years these has been much work characterising the
topology of real networks
\cite{Albert_data,Adamic,Faloutsos,Newman,Watts,Liljeros,Broder}.
This work has shown that our world is more complex than we had
originally imagined and has lead to the development of the idea of
a complex network. The most significant result arising from these
studies is that a power-law degree distribution appears to be very
common in real complex networks.

In this paper we examine empirically a variety of basic structural
and temporal properties of the internet discussion networks that
are created by internet users. The paper is organised as follows.
In the next section we introduce the different types of internet
discussions, and describe the scope of our empirical study. In
Section 3 we describe our results, both topological and temporal,
before summarising our findings in the final Section.

\section{Types of internet discussions}

Almost all internet discussions take place through the medium of
forums. Most internet information portals, on subjects such as
politics, accidents, sport, etc...,  include forums as part of
their web site. New topics are introduced to these forums on a
daily basis. Some portals give people fixed forums to discuss
common topics such as love, work and sport. Users cannot put
un-moderated messages into these forums; most forums have a person
or computer program - a \emph{moderator} - that acts as a referee
for the comments posted, and rejects posts that are deemed
unsuitable.

Another type of internet discussion are \emph{news groups}. These
are run on servers that normally contain an enormous number of
fixed topics. To become a member of a news group one needs a
computer program - a client of the news group. Nowadays all e-mail
programs like Microsoft Outlook or Mozilla Firefox have such a
client. The connection to the server is controlled by the
administrator of a server. Some servers are designed for anyone
and others themed for a group of people like students at a
university, employees of a company, etc. The administrator of a
server can block access to the server for people that break its
rules.

A third popular medium of internet discussion is a blog. There are
a number of websites where people can establish their own blog,
which usually takes the form of a diary of their day-to-day life.
Other people can discuss the blogs and express their opinions
about them to other readers or the owners of the blogs themselves.
The bloggers are usually able to place links to other blogs, which
are either on a topic related to their blogs or of general
interest to them, on their website. These links create a network
of blog owners \cite{makowiec}.

\subsection{Typical construction of internet discussions}

\begin{figure}
\begin{center}
\begin{tabular}{cc}
\includegraphics[width=4.5cm]{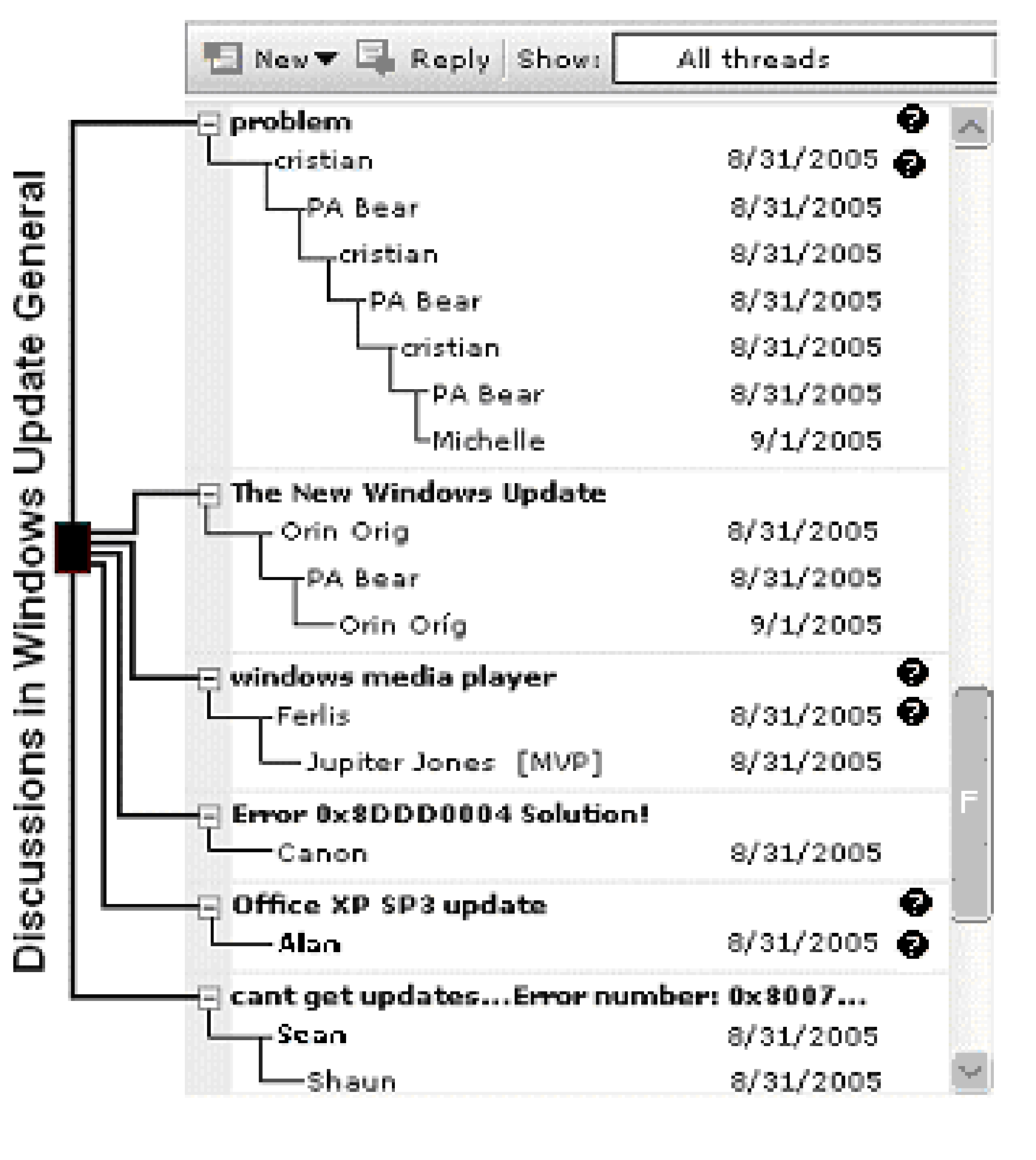} & \includegraphics[width=7.0cm]{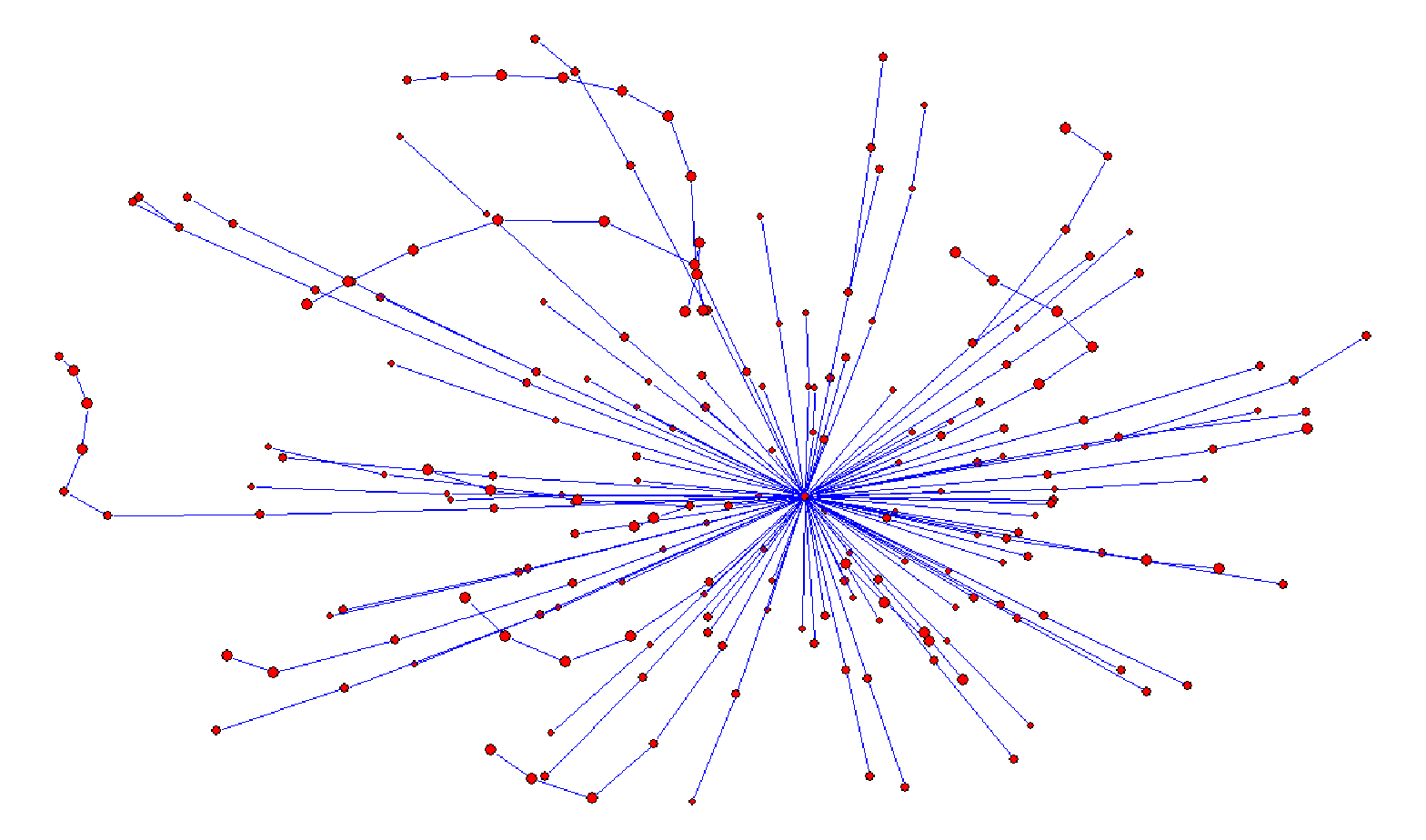}
\end{tabular}
\end{center}
\caption{(a) The typical structure of an internet discussion. The
black lines show links between messages and the responses to them.
(b) The tree-like structure of the small news group Physic,
$N=220$ nodes.} \label{news_exemp}
\end{figure}

For a news group and an internet forum the topic of the discussion
is a root node. The threads that initiate new discussions are
connected directly to the root node. When people contribute to a
forum they can either write a commentary on a previous opinion or
start a new thread. Every message is indexed by the name of the
author, its place in the hierarchy and its time of posting. In
this paper we treat each message as a node. We create a link
between a message and a responding or answering message. This
procedure creates a tree-like structure. Fig. \ref{news_exemp}
shows a typical structure of a small internet discussion.

We have investigated the network structure and temporal properties
of $3$ forums and $15$ news groups, whose data was collected from
two sources:
\begin{itemize}
  \item The internet forum on the web site www.onet.pl
  \item The news groups on the server news.student.pw.edu.pl
\end{itemize}

In the case of news groups the people who can contribute to a
discussion is limited by the fact that only computers inside the
university's network are allowed to login. Because of this only
students and academic staff have access to these discussions and
there are around $30,000$ of them each year. We did not measure
the number of active users of these news groups, but we suppose
that there are less than $5,000$.

The internet forum on www.onet.pl is part of the largest polish
news portal, which is used by around $50\%$ of all polish internet
users.

Almost all internet discussions that we have collected, were
created at different time. However for internet forums the period
of collected data is between $2001-2005$ and for news groups the
period is $2002-2005$.

\section{Empirical results}

We study empirically a number of properties of real internet
discussions. Our networks are trees and consist of messages, not
users, so we are unable to study properties such as the clustering
coefficient or to define communities. Similarly it would be
fruitless to study node mixing or the betweenness centrality,
which were studied in \cite{makowiec,Zhongbao,Goh,Valverde}. Thus,
the structural properties we examine are the degree distribution,
the average and maximal distance, the distribution of distances
between messages and their root nodes and the average supremacy
\cite{Holyst} of each node as a function of degree. The temporal
properties we examine are the distribution of time between a
message being posted and there being a response to it, the
activity time series and its correlation function. With the
temporal properties we distinguish between network time; time in
which one message is posted in one time step and message $i$ is
added at time $i$, and real time; the actual time that messages
were posted in our experimental data. Where appropriate we present
results for both an internet forum and a news group, for the
largest and most representative examples.

\subsection{Degree distribution}

All the networks we examined were found to have power law degree
distributions

\begin{equation}
p(k) \sim k^{-\gamma}.
\end{equation}

Table \ref{tab_exep} lists the topics of these discussions, their
size $N$, the exponent $\gamma$ of their power law degree
distribution, the maximal distance $R_{max}$ from the root node,
the ratio of the number of threads $n_1$ over the total number of
messages $N$ and finally the average distance of all nodes in the
network from the root node $<r>$.

The internet forums generally have a lower exponent $\gamma$  than
the news groups. In particular, the exponents for forums are in
the range $3.28 < \gamma < 4.24$ and for news groups $3.90 <
\gamma < 5.84$.

\begin{figure}
\centering \subfigure {\psfig{figure=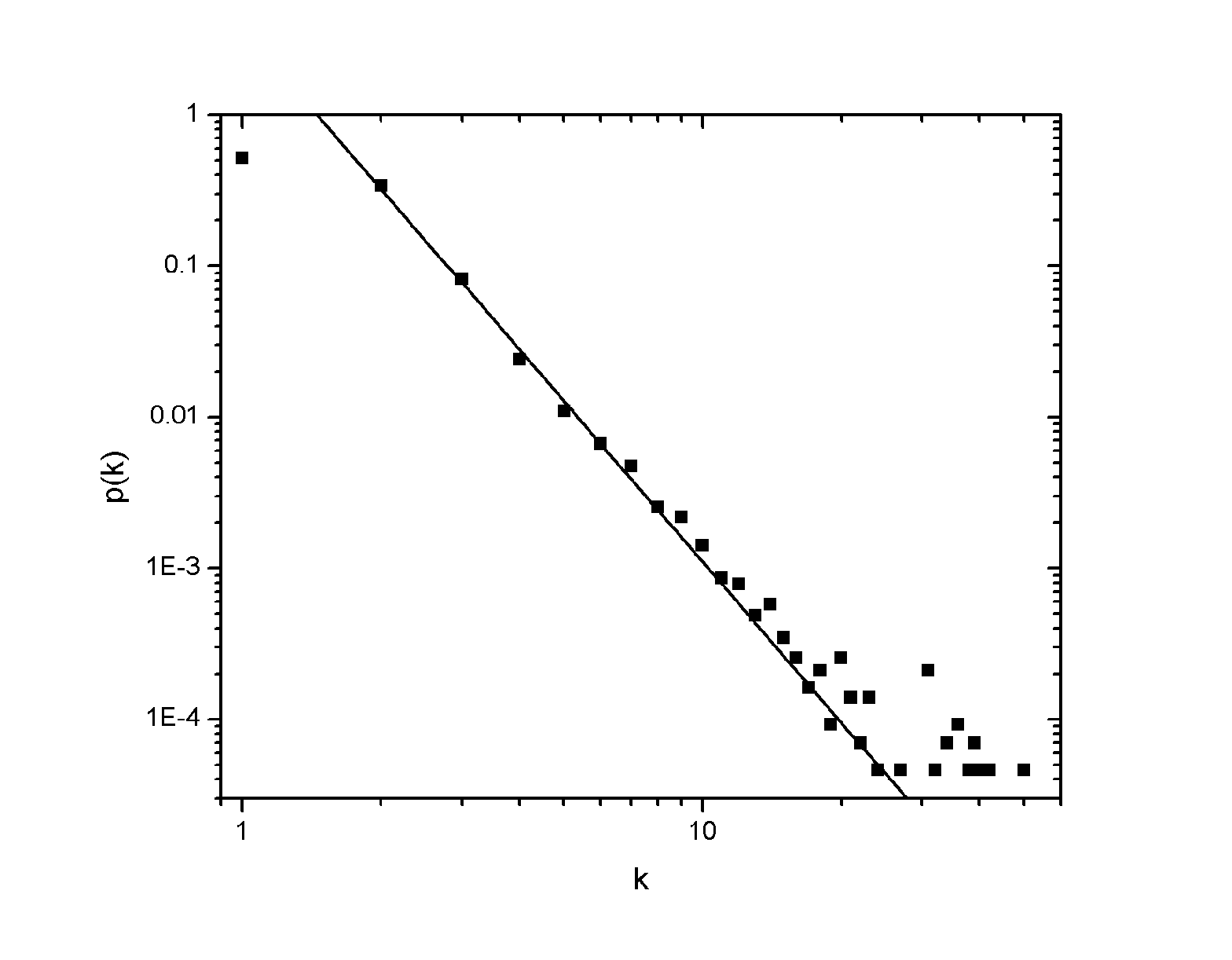,height=4.2cm}}
\subfigure {\psfig{figure=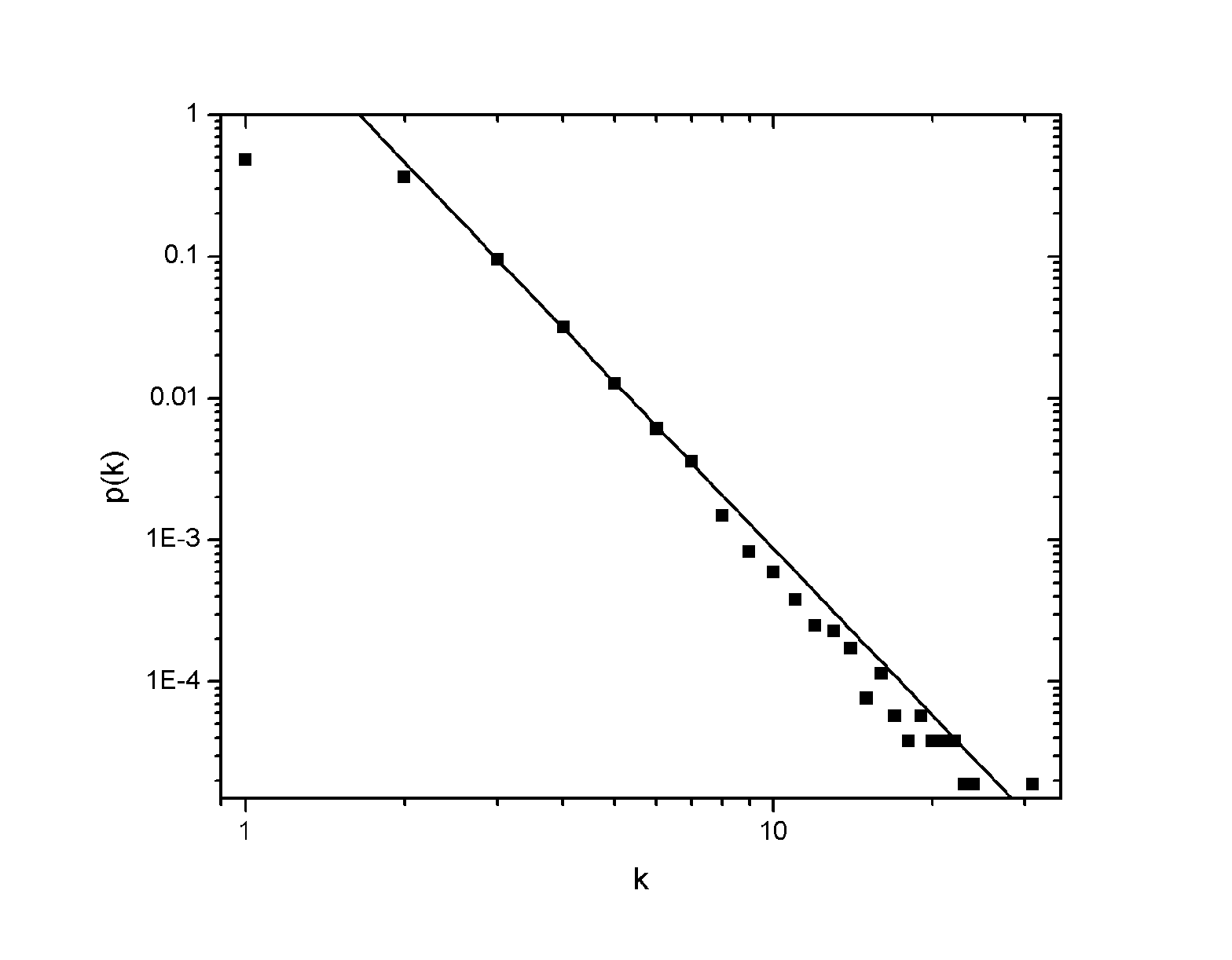,height=4.2cm}} \caption{The
degree distribution for the internet forum Poland in the EU (a)
and the news group Humor (b). The exponents $\gamma$ are
$\gamma=3.53$ for forum Poland in the EU (a) and $\gamma=3.90$ for
news group Humor (b).} \label{deg_dist}
\end{figure}

Fig. \ref{deg_dist} shows a typical degree distribution for the
forums and the news groups. The networks have few nodes with high
degree, even for the larger networks, with only 7 networks having
a maximum degree $k_{max}>30$. For the news groups the largest
degree is around 20.

In all networks we examined the number of nodes with degree $1$
was similar to the number of nodes with degree $2$, that is
$p(1)\approx p(2)$. This seems to be because people like to argue
and preferentially create chain structures in threads, and also
because people also sometimes respond to their own messages. This
behaviour creates more nodes with degrees $k=2$, $k=3$, etc... and
shifts the degree distribution towards higher values of $k$.

\subsection{Time interval distribution $T(\tau)$}

Internet users visit news portals to update themselves on the
recent news, and some of them will discuss this news in a forum.
In most cases they will only discuss the very latest news, and
only very interesting topics will be discussed by users over a
long period of time. The same rule applies for messages, only
interesting or very controversial opinions are discussed for a
long time period. This is why messages age very quickly and are
soon forgotten. The influence of aging is the reason for the large
exponent $\gamma$ in these networks and for the lack of nodes with
large degree.

There have been a number of attempts to model the effect of aging,
see for instance, \cite{Amaral,Dorogovtsev,DorogovtsevS,Zhu}. The
fundamental quantity in these models is $\pi(k,t,\tau)$, the rate
of attaching a new node to a node of degree $k$ and age $\tau$ at
time $t$. All these models assume that $\pi(k,t,\tau)$ a separable
function of the degree and the age of the node. In particular,
Dorogovtsev and Mendes \cite{Dorogovtsev,DorogovtsevS} modelled
this aging by assuming that incoming nodes are linked to a node
with degree $k$ and age $\tau$ with rate $\pi(k,t,\tau)=A(\tau)k$,
where $A(\tau)$ is some aging function, given by

\begin{equation}
 A(\tau)\sim \tau^{-\beta}.
\label{equ1}
\end{equation}

They found that the degree distribution of this network remained
power law, $p(k)\sim k^{-\gamma}$ in the large time limit but with
an exponent $\gamma$ that strongly depends on the exponent $\beta$
in the aging function \cite{Dorogovtsev}.

Unfortunately, $A(\tau)$ is not easily measured empirically, as
attempts to verify that some real networks were grown by
preferential attachment without aging clearly illustrate
\cite{new}. Instead, we have measured a related quantity,
\cite{hajra}, the time interval distribution. This is the
distribution of times between a message and a response, for all
the internet discussions. More precisely, where a message $j$,
posted at real time $t_j$, receives a response $i$ at real time
$t_i$, we have studied both the distribution of the real time
interval $\tau=t_i-t_j$, $T(\tau)$ and the distribution of network
time interval $i-j$. The distribution $T(\tau)$ is related to the
degree distribution at time $t$, $p(k,t)$ via

\begin{equation}
T(\tau)=\int{w(k,t,\tau)p(k,t)dk dt}
\end{equation}

where $w(k,t,\tau)$ is the probability that a node of degree $k$
at time $t$ waits another $\tau$ timesteps before gaining an edge.
This latter function contains, implicitly, two temporal processes,
the natural waiting time for a new edge which exists in all
growing network models, plus the effect of the aging identified
and modelled in \cite{Amaral,Dorogovtsev,DorogovtsevS,Zhu}.
However, for $1<<\tau<<t$, we expect that the effect of the former
will be exponential in $\tau$ on $T(\tau)$ whereas if there is
appreciable aging, this will manifest itself as a fat tail in
$T(\tau)$ for large $\tau$.

\begin{figure}
\centering \subfigure {\psfig{figure=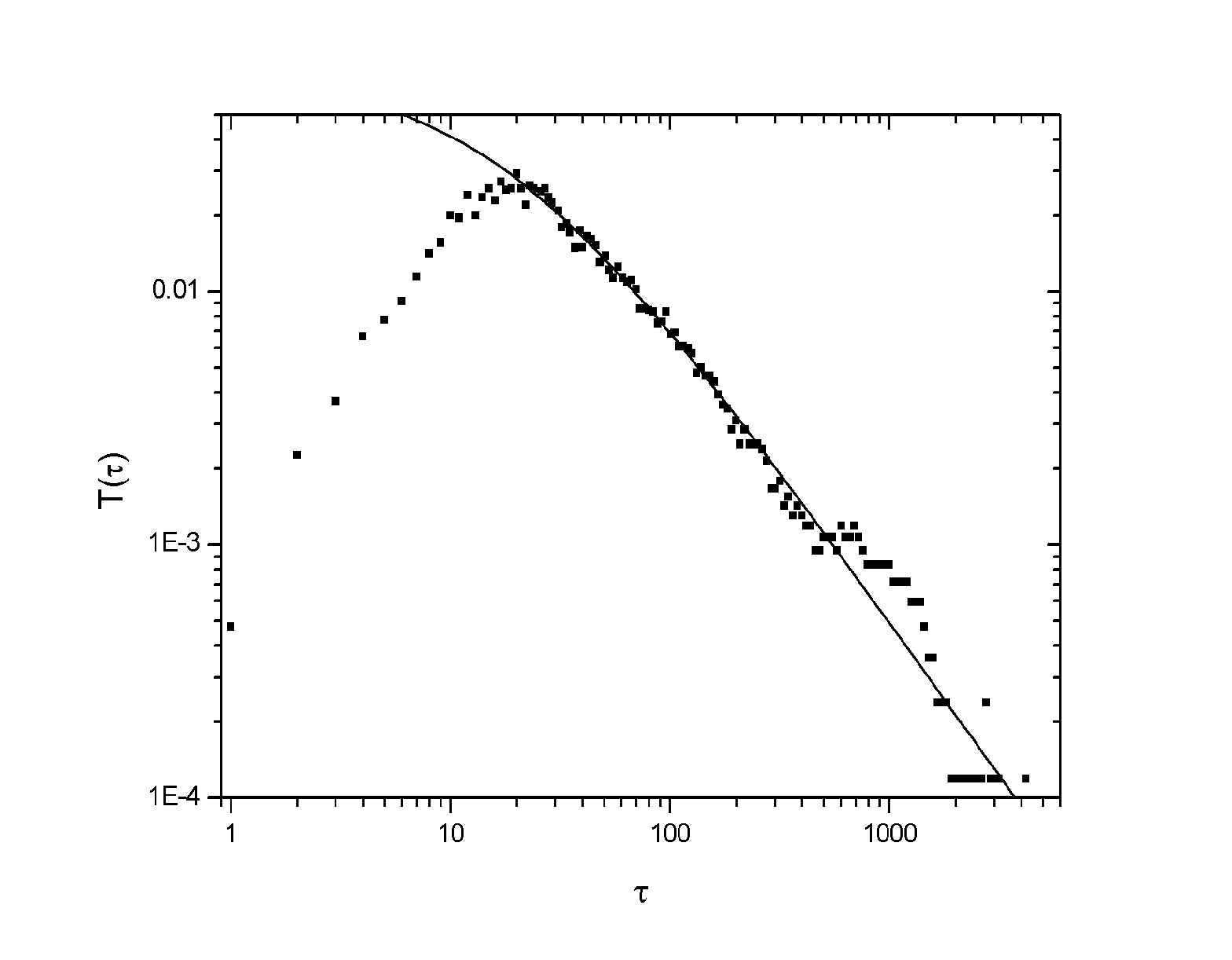,height=4.0cm}}
\subfigure {\psfig{figure=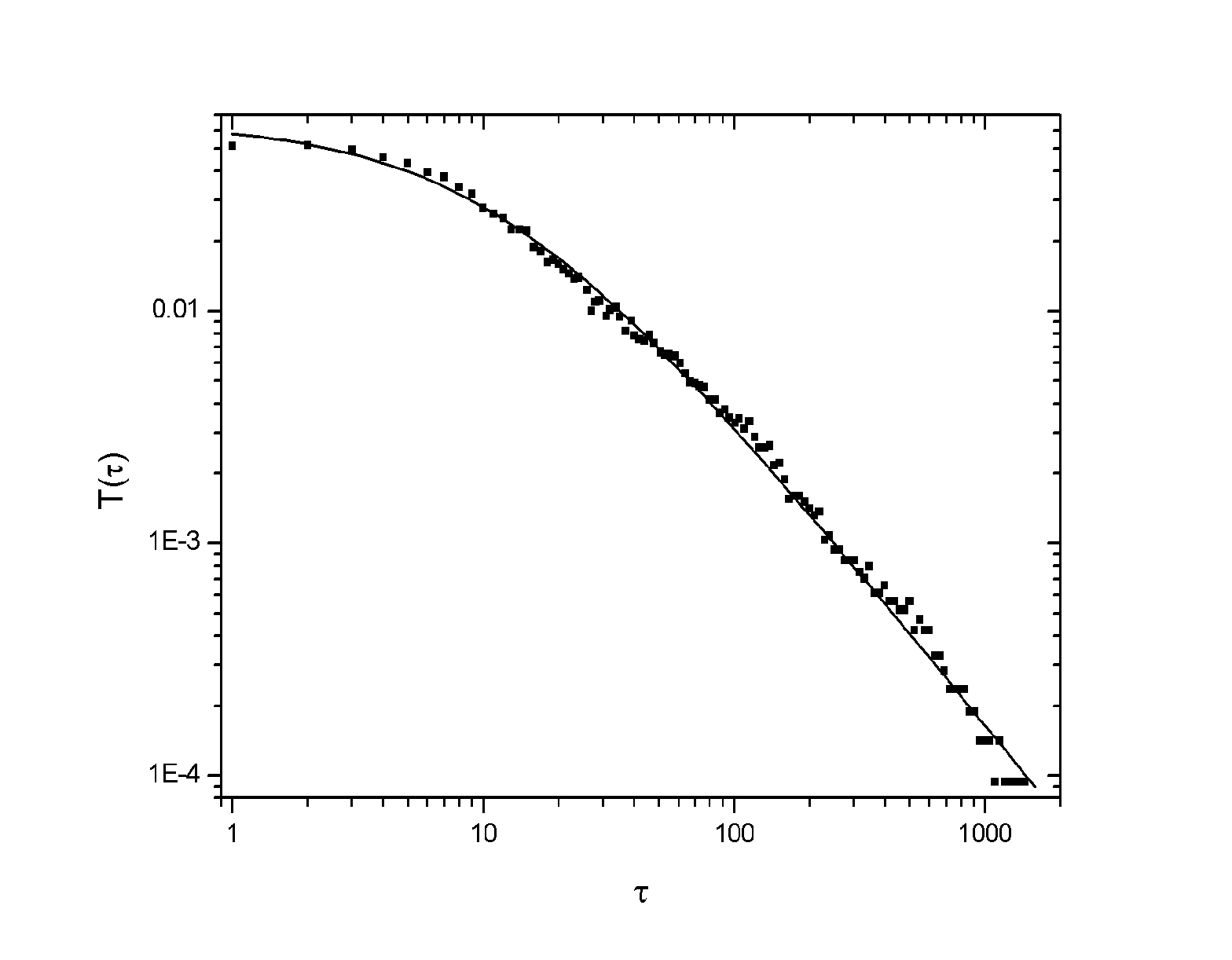,height=4.0cm}} \caption{The
time interval distribution in real time for (a) the forum Poland
in the EU and (b) the news group Humor. The exponent $\delta=1.25$
for (a) and $\delta=1.33$ for (b). The real time unit is 1
minute.} \label{TIM_dist}
\end{figure}

In fact our results for real time show that in an internet news
group messages age and have a power law $T(\tau)$. In Fig.
\ref{TIM_dist} we show that the time interval distribution

\begin{equation}
  T(\tau)\sim[\tau+\tau_0]^{-\delta}.
\label{dorogov}
\end{equation}

The positive slope of the curve in Fig. \ref{TIM_dist}a for small
time intervals results from the presence of the \emph{moderator}
in the forum www.onet.pl. The moderator has to check each message
and this takes some time. Fig. \ref{TIM_dist}b shows that Eq.
(\ref{dorogov}) gives a good approximation to the empirical
measurements.

\begin{figure}

\centering \subfigure {\psfig{figure=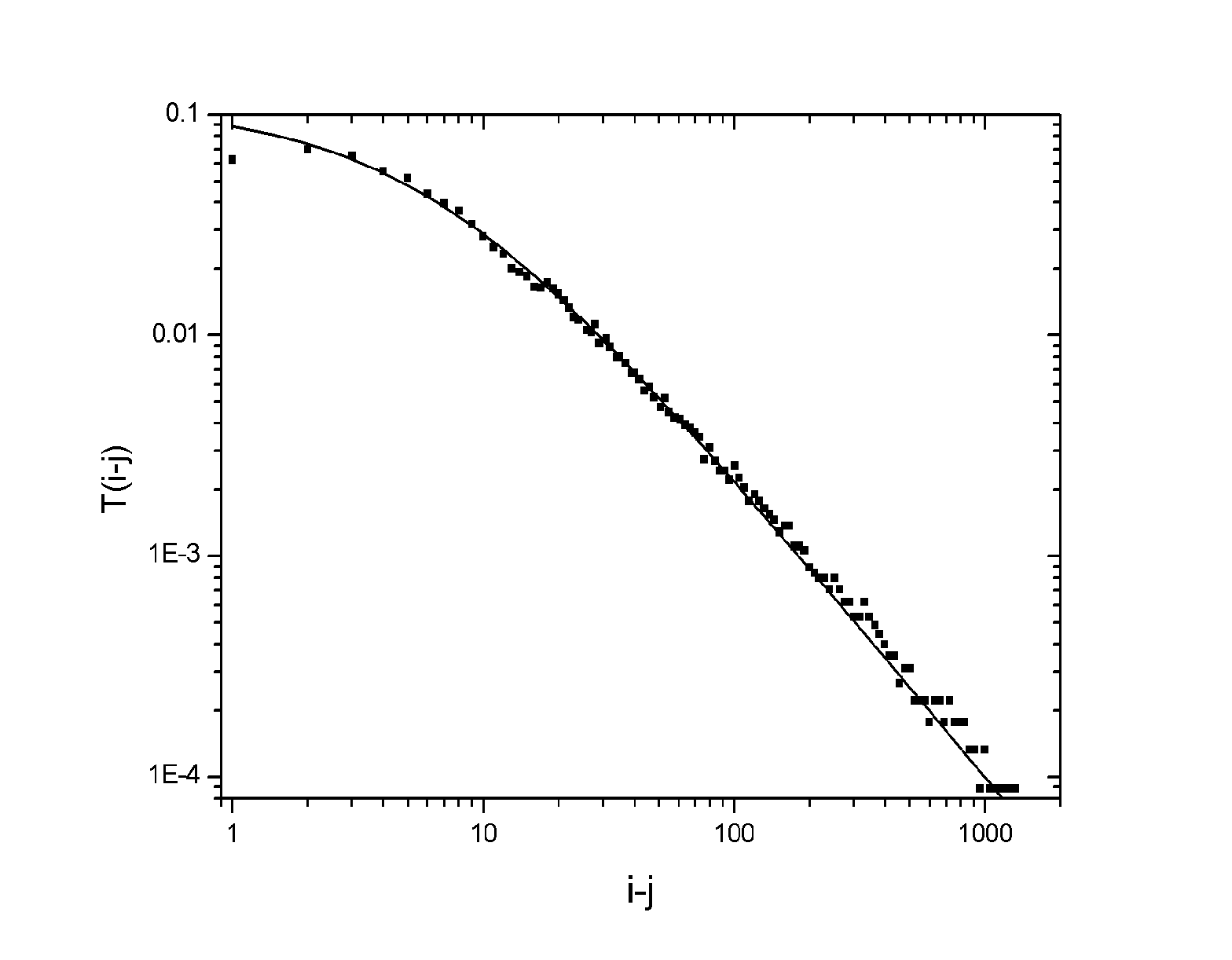,height=4.0cm}}
\subfigure {\psfig{figure=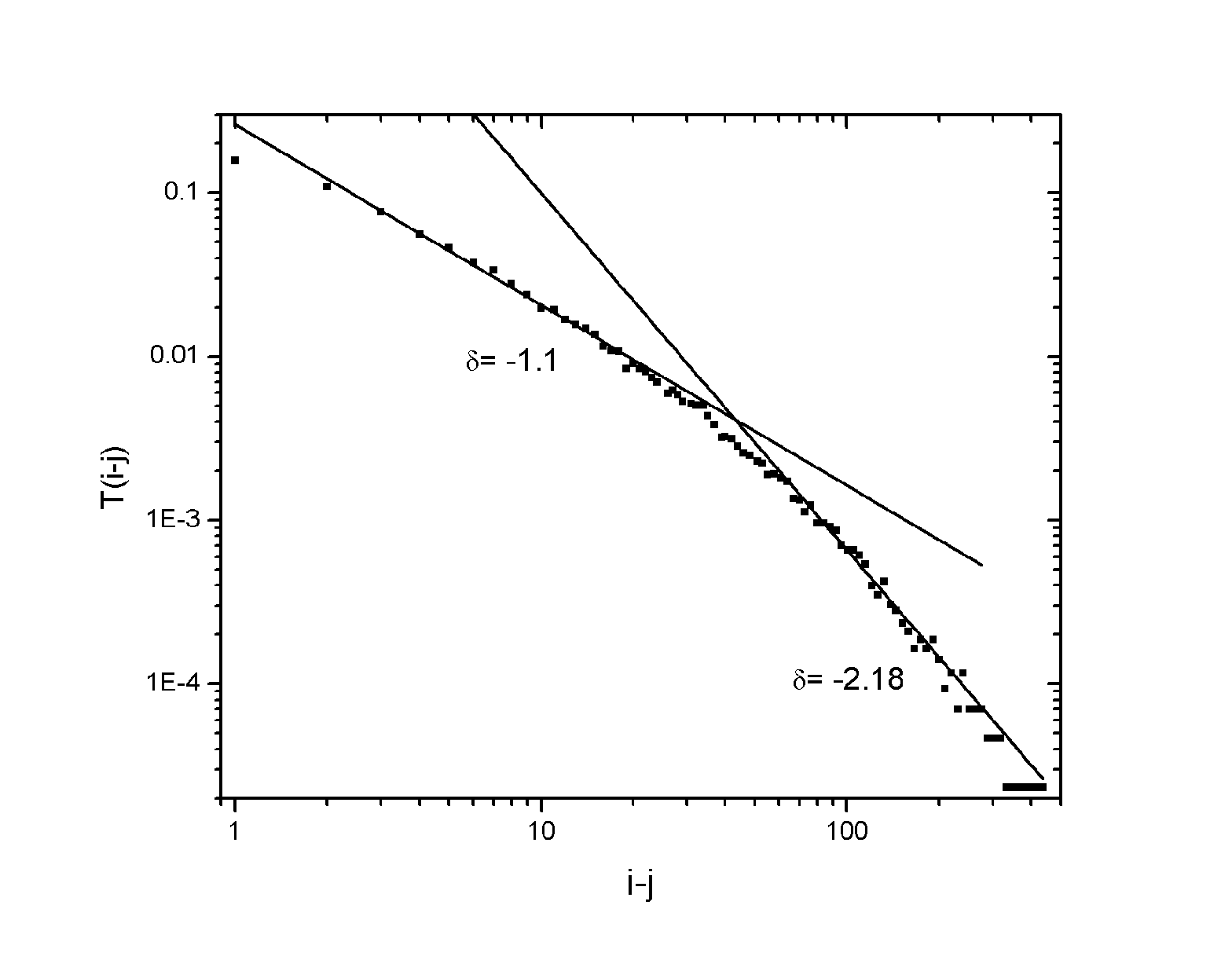,height=4.0cm}}
\caption{The time interval distribution in network time for (a)
the forum Poland in the EU and (b) the news group Humor. The shape
of Fig. (a) follows Eq. (\ref{dorogov}) with $\delta=1.37$. Fig.
(b) is described by composite power laws with exponents
$\delta=1.1$ and $\delta=2.18$.} \label{TIM_dist2}
\end{figure}

In Fig. \ref{TIM_dist2} we show the time interval distribution in
network time and this merits two observations. Firstly, there is a
change in the time interval distribution. For all news groups (but
not for the forums) we obtained time interval distributions with
two regimes of aging. For each news group there is a
characteristic, cross-over time interval $t_c$ after which
messages start aging faster. This characteristic time is different
for each network.

Secondly, the shape of time interval distribution for internet
forum is not effected by a \emph{moderator} and exactly follows
Eq. (\ref{dorogov}). This means that for small time intervals
messages age slower and for large intervals faster but the change
is smooth and without the critical point observed in news groups.

The power law behaviour of the time interval distribution was
studied by Barab\'{a}si \cite{Barabasi} for an e-mail exchange
group. By simulating the types of activity of internet users, it
was shown that only the \emph{burst} activity results in power law
distributions, $A(\tau)\sim \tau^{-\delta}$, where $\delta=1$.
Fig. \ref{TIM_dist2}b shows that for small network time intervals
the index $\delta$ is close to $1$. For all news groups
$\delta\in(1.0, 1.5)$. Because of the \emph{moderator} the results
for internet forums are disturbed, however the value of
$\delta=1.37$ is still close to $1$ (Fig. \ref{TIM_dist2}a).

\begin{figure}
\centering \subfigure {\psfig{figure=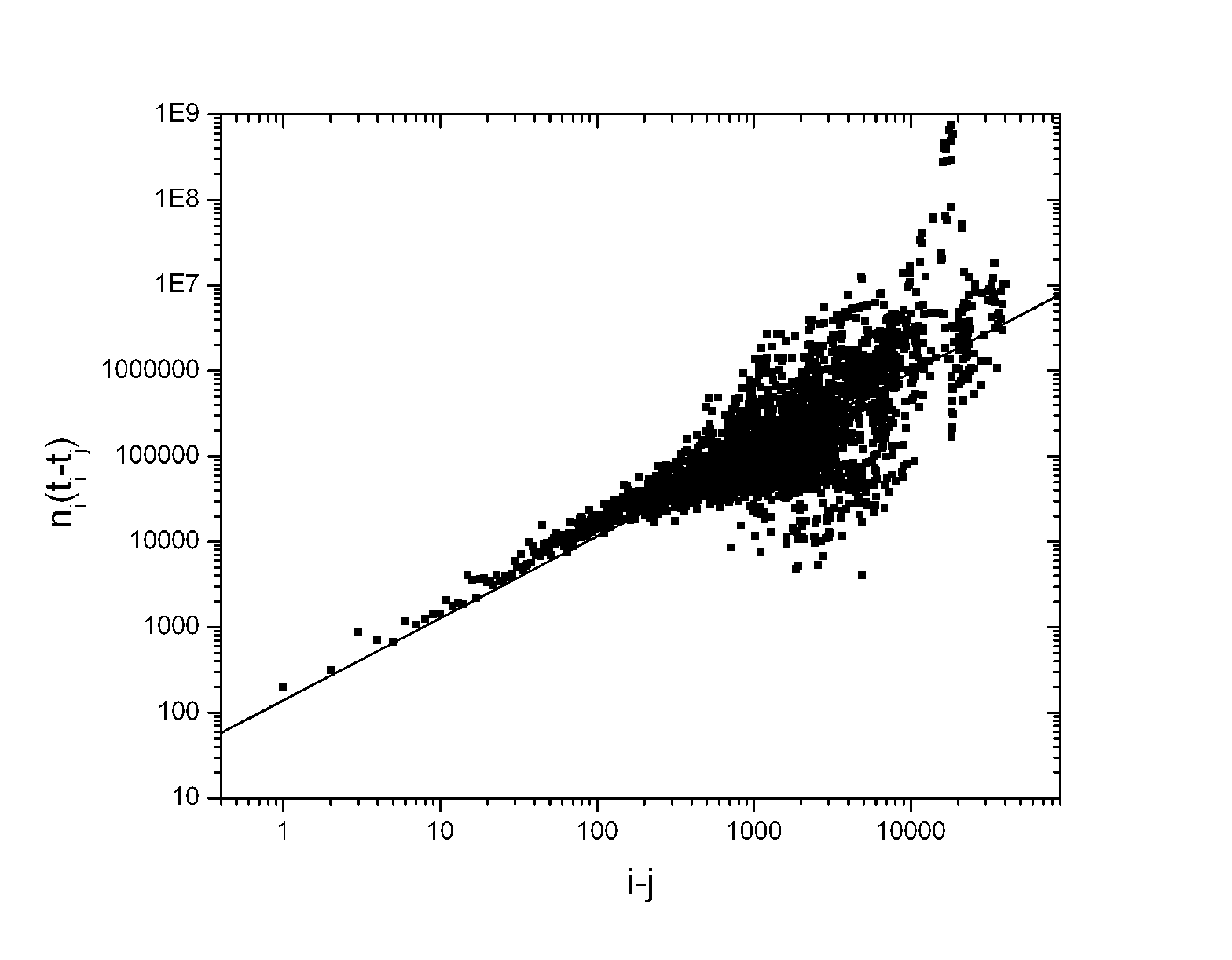,height=4.0cm}}
\subfigure {\psfig{figure=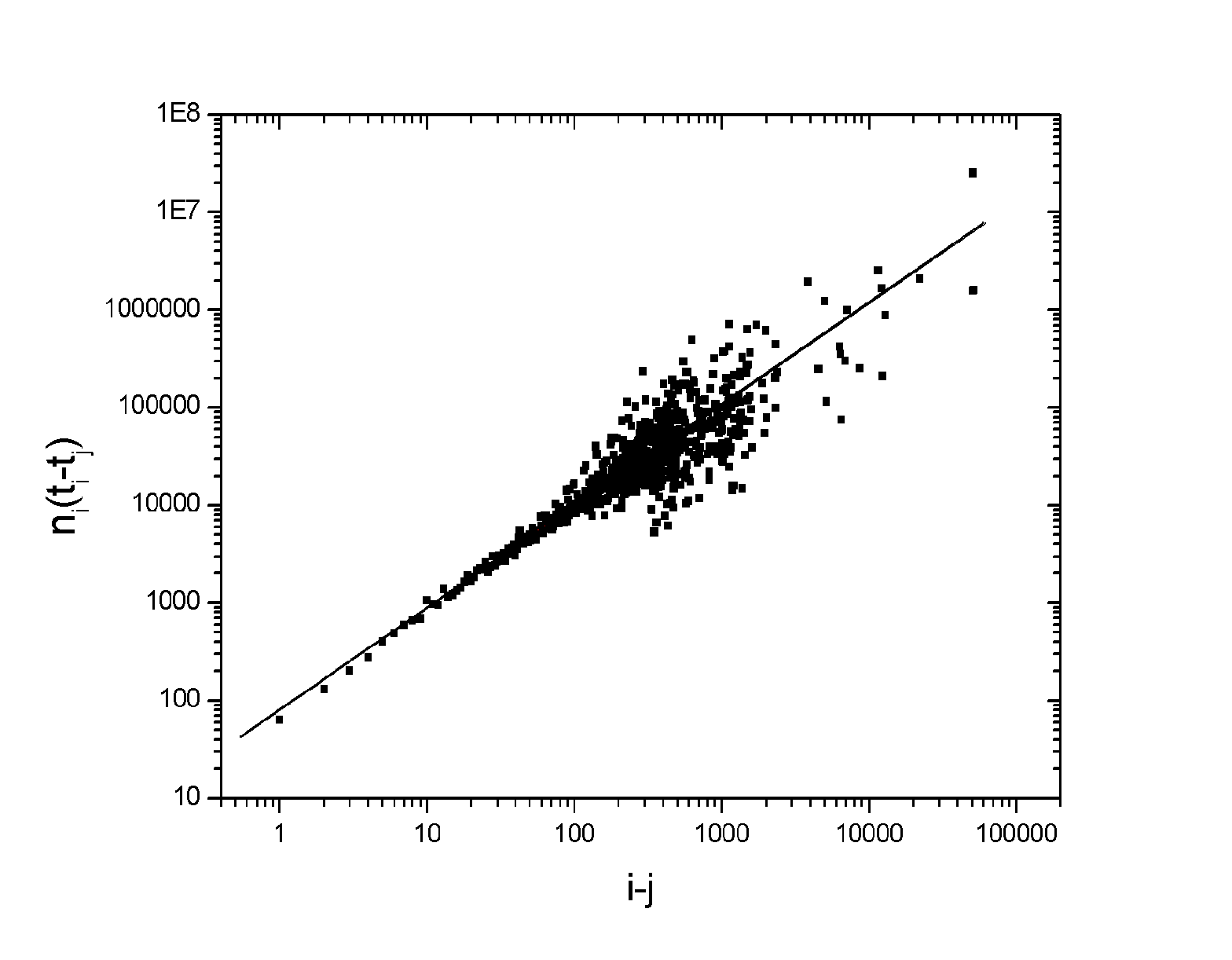,height=4.0cm}}
\caption{The average value of the real time interval multiplied by
the activity as a function of the network time interval for (a)
the forum Poland in the EU and (b) the news group Humor.}
\label{i-r-time}
\end{figure}

We also studied the relationship between the network time interval
and the real time interval. Of course these are related by the
fact that the activity $n(t_i)$, which is the number of messages
that were posted in time $t$ satisfying $t_i<t<t_{i+1}$, can be
approximated by $n(t_i) \approx (i-j)/(t_i-t_j)$. Our empirical
results show that, as would be expected, on average the relation
is linear with

\begin{equation}
n(t_i)(t_i-t_j) \sim \epsilon(i-j)
\end{equation}
with $\epsilon = 1.04 \pm 0.02$ for the internet forum Poland in
the EU Fig.\ref{i-r-time}a and $\epsilon = 0.96 \pm 0.02$ for the
news group Humour Fig.\ref{i-r-time}b.

\subsection{Activity}

\begin{figure}
\centering \subfigure
{\psfig{figure=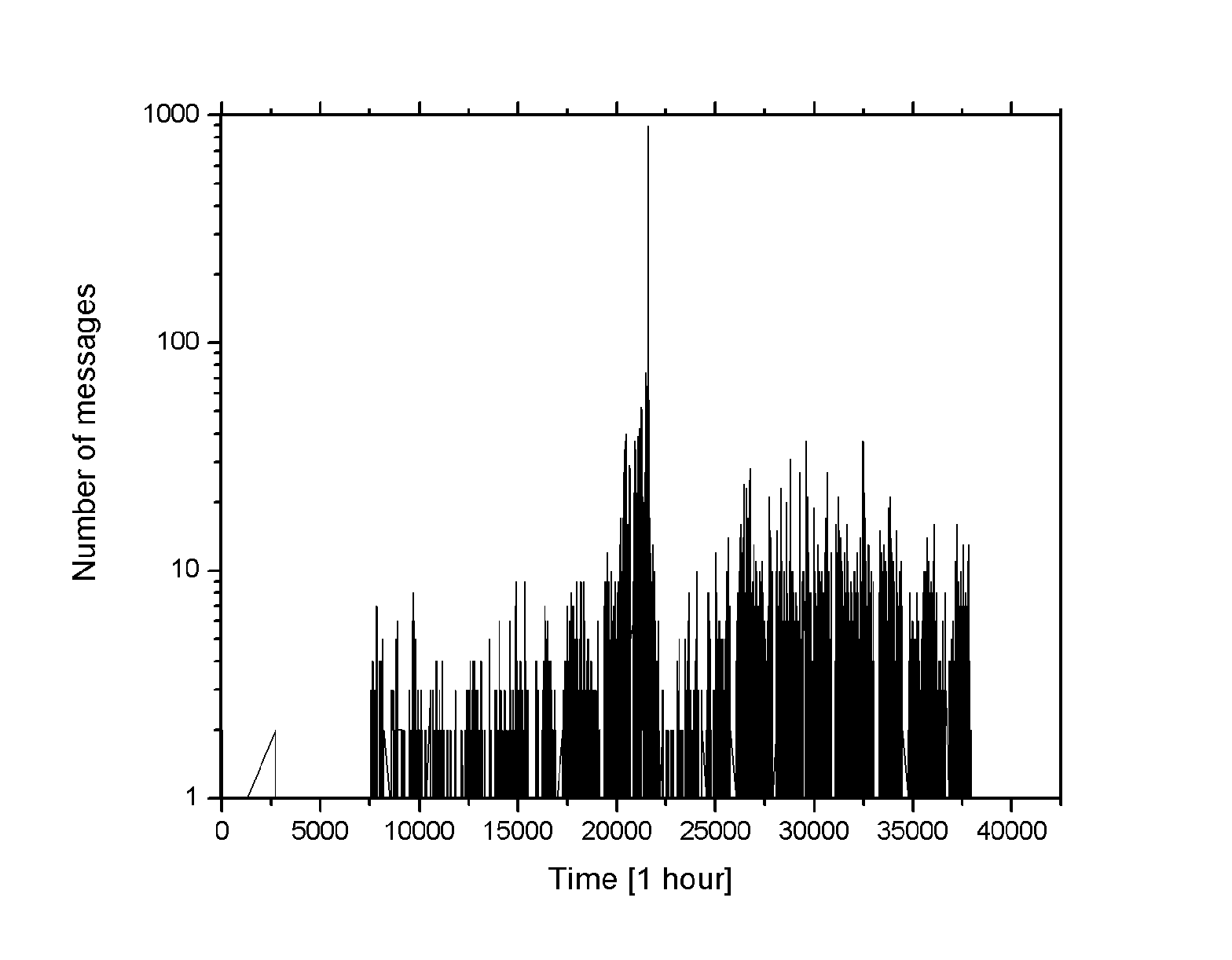,height=4.0cm}} \subfigure
{\psfig{figure=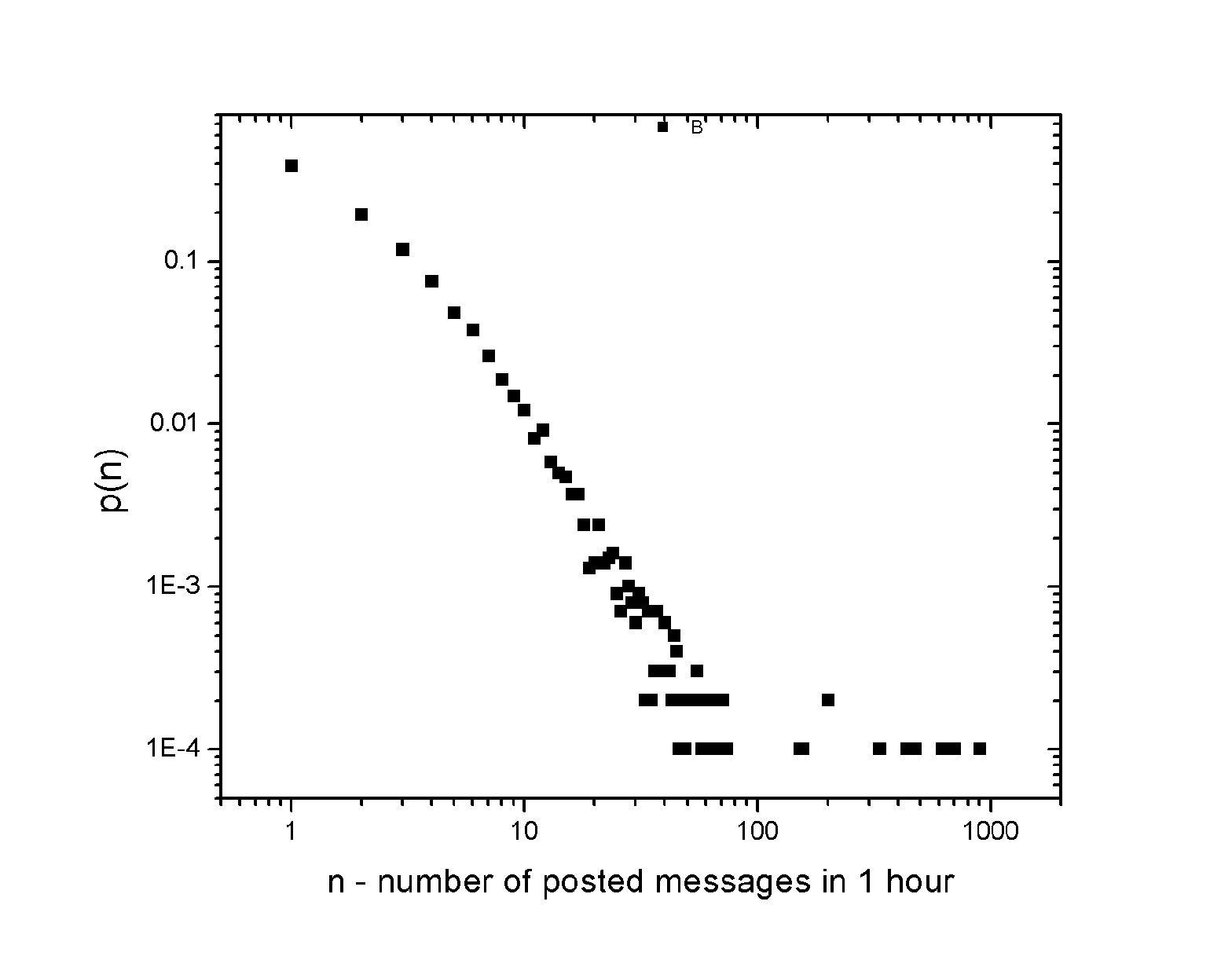,height=4.0cm}} \caption{The activity
time series (a) and the activity distribution function (b) for the
forum Poland in the EU.} \label{activity}
\end{figure}

We define the activity of a news group as the number of messages
posted in a given time interval. In Fig. \ref{activity} we show
the activity time series and the distribution of activity for the
discussion forum Poland in the EU. Here we have measured the
number of messages posted in one hour. As one can see, there is a
variation of activity over a wide range of scales.

The peak, in which 898 messages were posted in a single hour,
corresponds to the time when Poland was voting in the accession
referendum to the EU. This hiatus can be seen in the activity
distribution, corresponding to the points to the right in Fig.
\ref{activity}b, away from the main curve. We examined the
distribution of activity for all our news groups, and found that
all the distributions were fat-tailed, with distributions that
ranged from power-law to Kohlrausch, $\sim exp(-\tau^a)$, with $0
< a < 1$.

We have measured the correlation function $C(\tau^*)$ of the
activity time series, $n(t)$, defined by

\begin{equation}
C(\tau^*)=\displaystyle\sum_{i=0}^{i_{max}}[n(t_i)-<n>][n(t_i+\tau^*)-<n>]
\end{equation}
where $t_i=t_0i$ and $<n>$ is the mean number of messages posted
per time $t_0$ over the whole time series. We studied $t_0=1 hour$
and $t_0=1 day$.

\begin{figure}
\centering \subfigure
{\psfig{figure=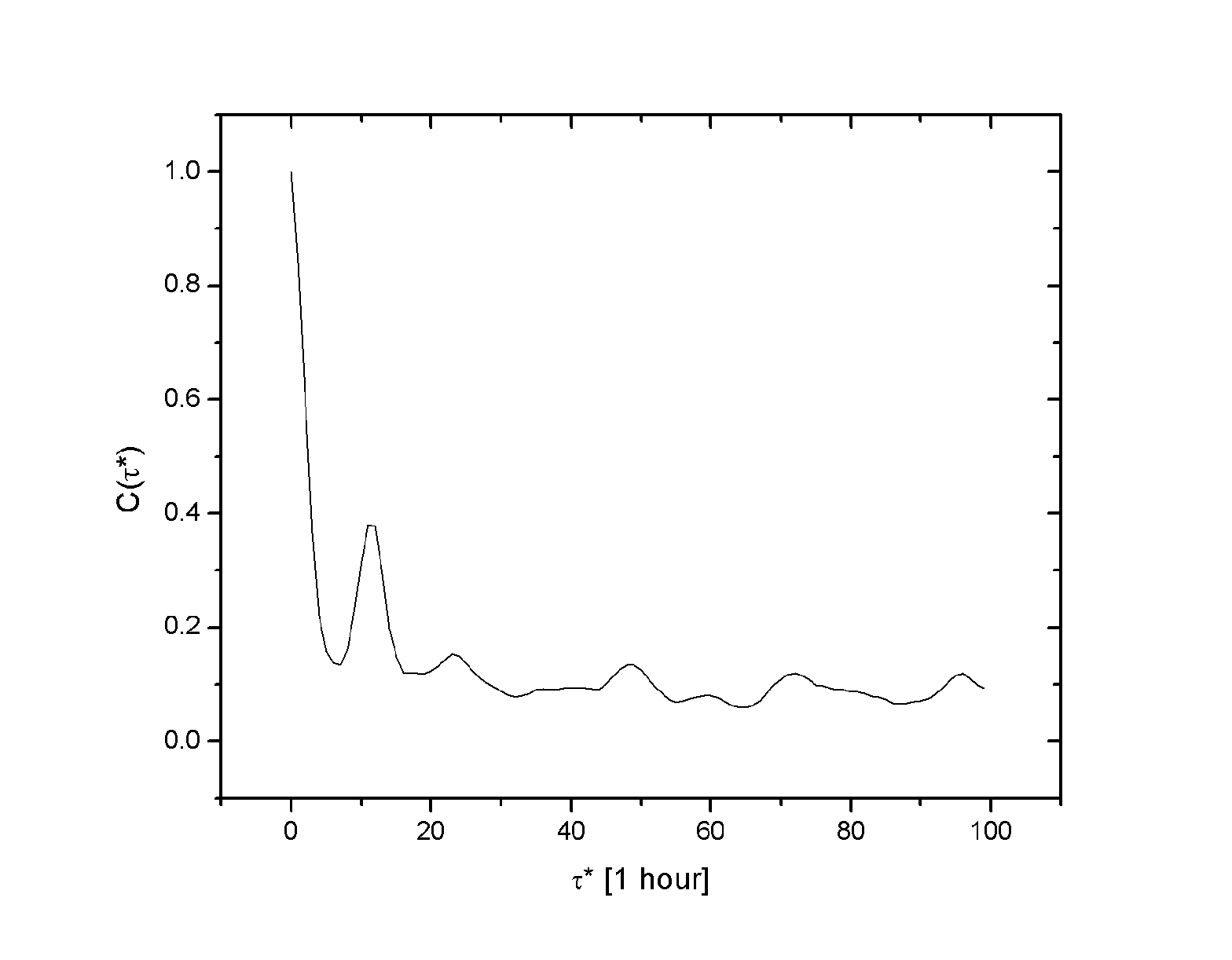,height=4.0cm}} \subfigure
{\psfig{figure=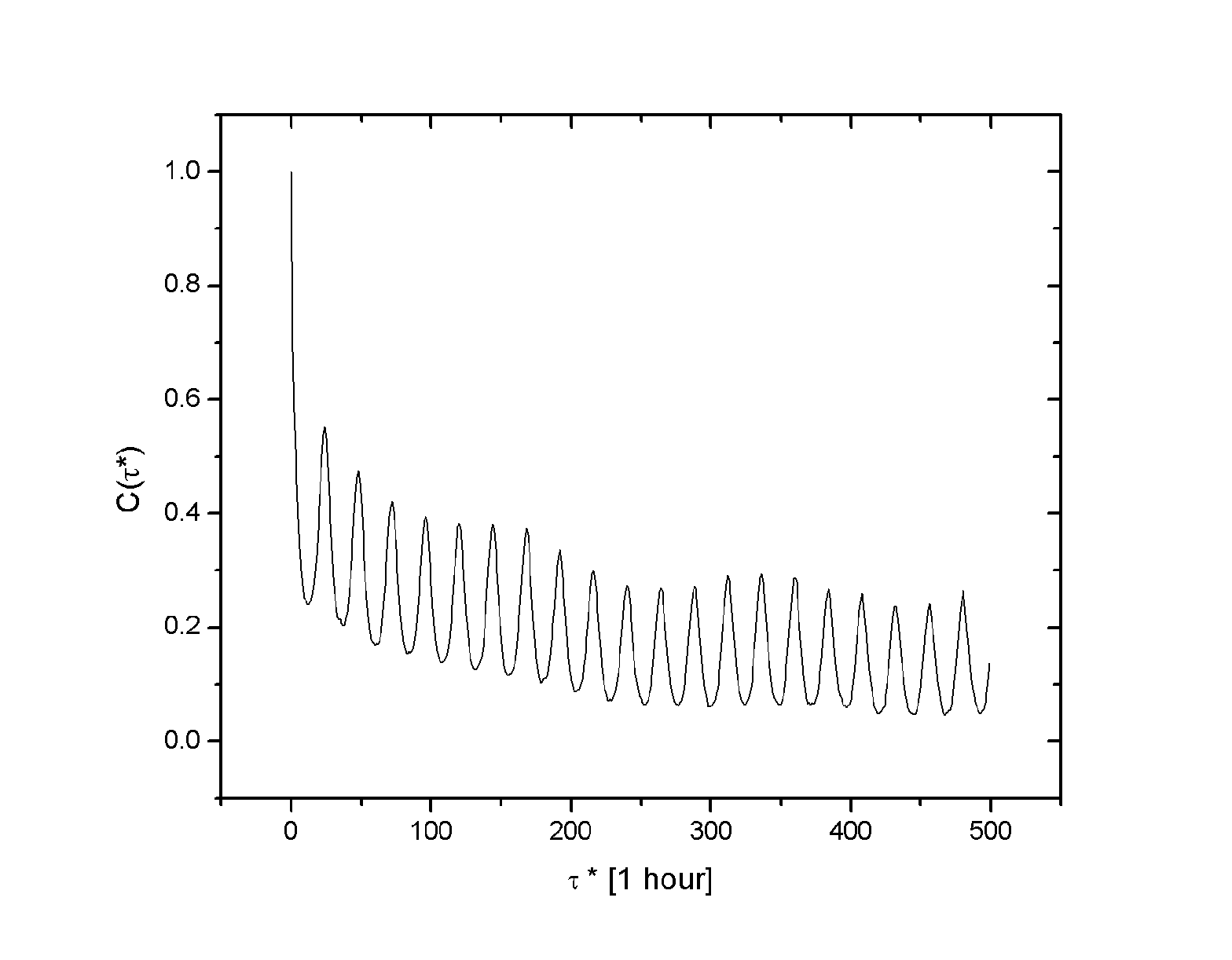,height=4.0cm}} \caption{The
correlation function $C(\tau^*)$ for (a) the forum Poland in the
EU and (b) the news group Humor, with a time step $t_0=1$ hour.}
\label{corel1}
\end{figure}

All the internet discussions indicate a correlation for $\tau^*=24
$ hours, which shows the daily routine of the internet discussion
users (see for instance Fig. \ref{corel1}b). We also found a weak
correlation for news groups on the time scale of one week, which
is probably connected to the higher activity over a weekend. This
is somewhat less pronounced, as Fig. \ref{corel1}a illustrates.
Some news groups also show correlations for very long times, for
instance for $\tau^*$ equal to 180, 270 and 365 days. These were
seen in news groups that are only used by students and these long
correlations are connected the academic holiday and semester
structure. There is an interesting correlation for $\tau^*=12$
hours in the forum Poland in the EU. This correlation is generated
by the before-after work activity of the discussion users.

\subsection{The distance distribution $D(r)$}

\begin{figure}
\centering \subfigure {\psfig{figure=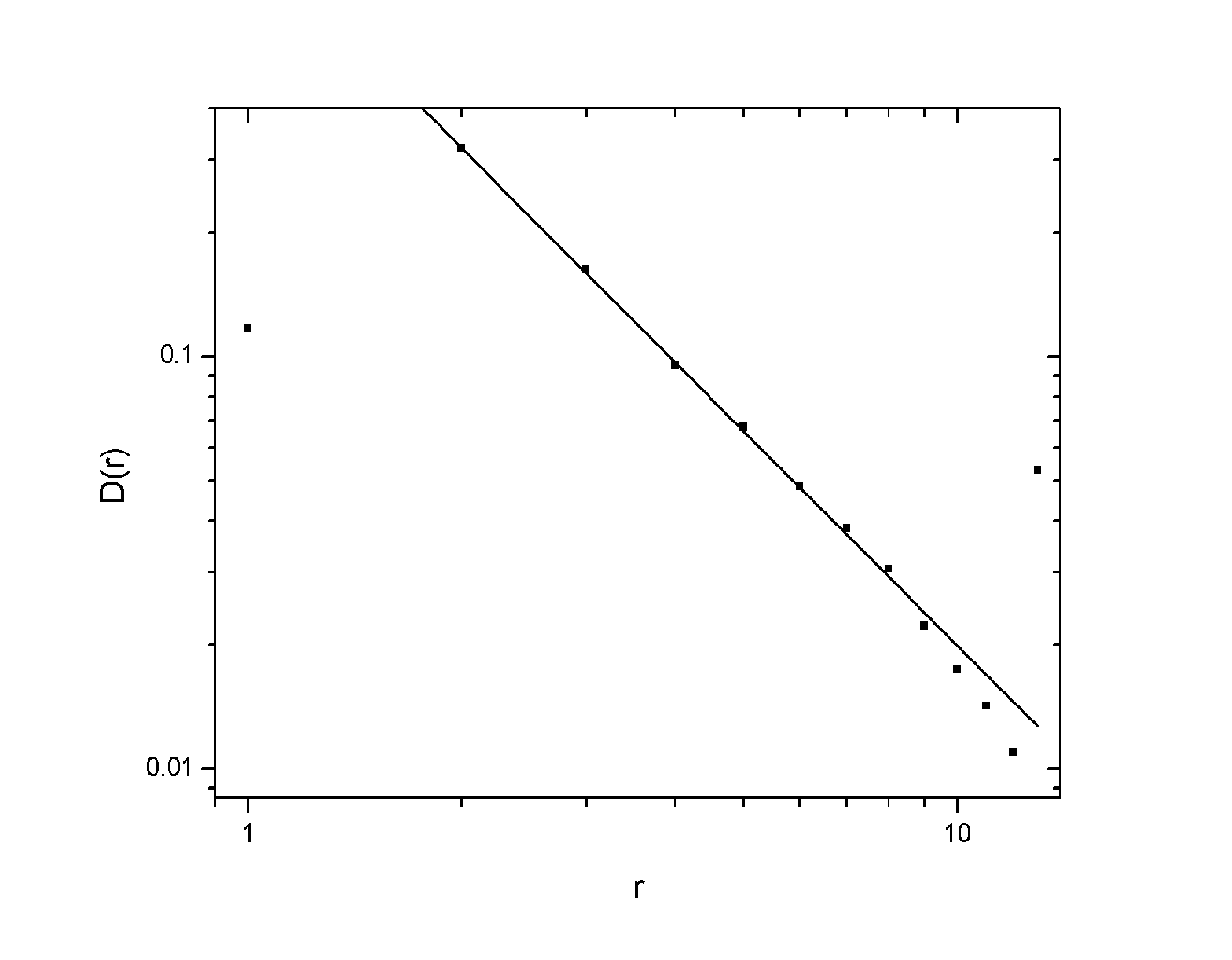,height=4.0cm}}
\subfigure {\psfig{figure=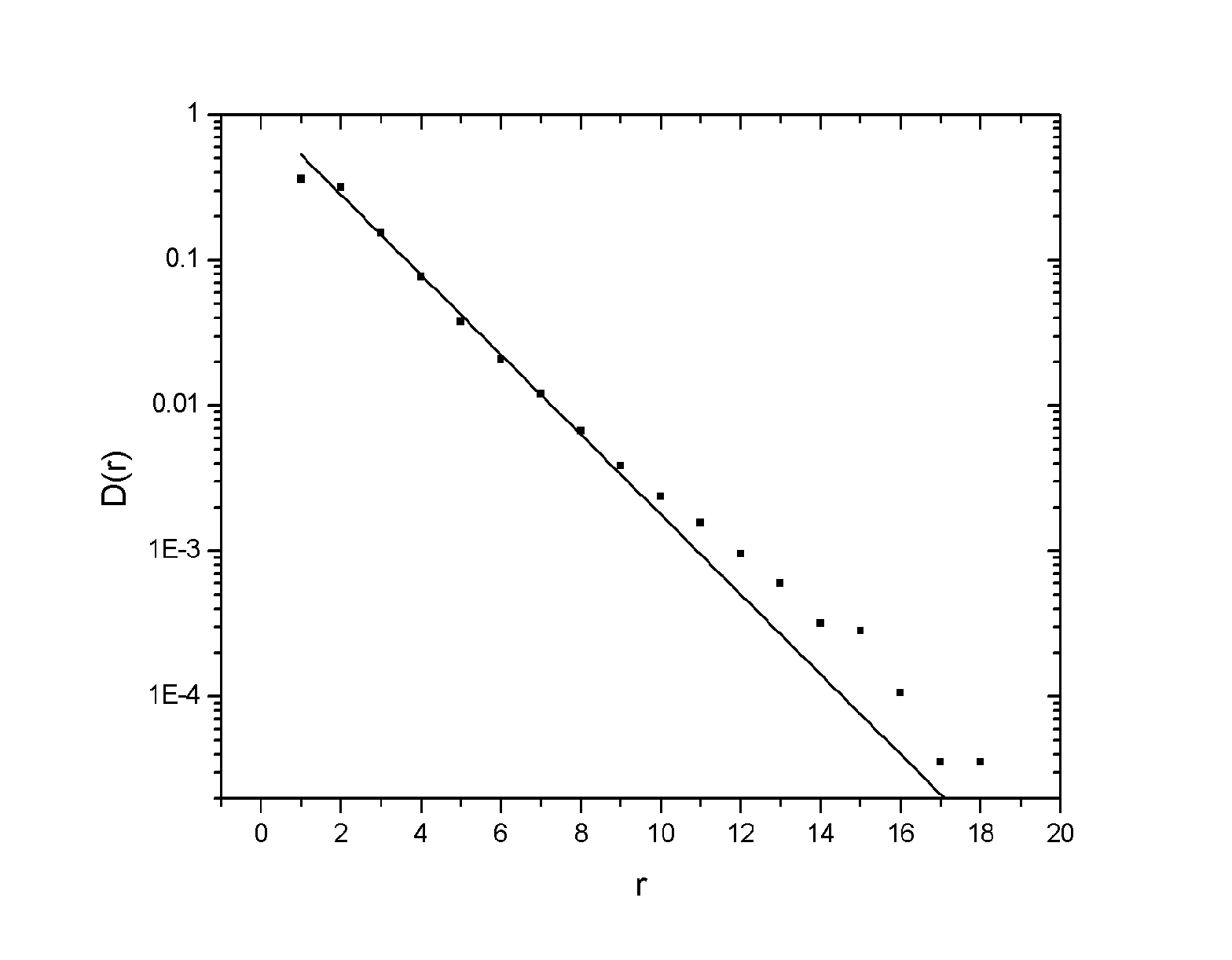,height=4.0cm}}
\caption{The distance distribution $D(r)$ for (a) the forum Poland
in the EU and (b) the news group Electronics. The distribution for
the forum has a power law behaviour $D(r)\sim r^{-\upsilon}$ with
exponent $\upsilon=1.73$ and the distribution for news group has
an exponential behaviour.} \label{D(r)}
\end{figure}

$D(r)$ is the distribution of the number of edges, between each
node in the network and its root node. For all the networks the
maximum distances are small. Almost all the news groups exhibit an
exponential $D(r)$, such as that illustrated for the news group
Electronics in Fig. \ref{D(r)}b. Of the news groups, only Humor
has a distance distribution close to a power law.

The distance distributions for forums are modified by the software
used to manage the forum, which only allows a maximum distance of
$r=13$. A message that somebody wants to post to a message with
$r=13$ is added to previous the message with $r=12$. This results
in the large value of $D(13)$ seen in Fig \ref{D(r)}a.
Nevertheless, despite this limitation the distance distributions
can show power law behaviour, as Fig. \ref{D(r)}a illustrates.

\begin{figure}[htb]
\begin{center}
\includegraphics[width=6.0cm]{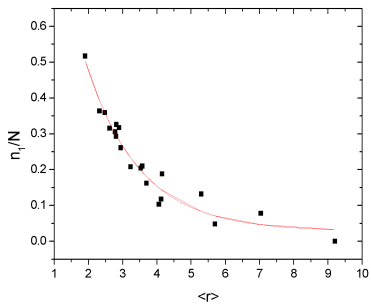}
\end{center}
\caption{The ratio of the number of threads $n_1$ to the total
number of messages $N$ as a function of the average distance from
the root $<r>$. The curve is fitted an exponential function
$f(<r>)\sim e^{-<r>/<r_0>}$, where $<r_0>\approx 1.58$. }
\label{avR_n1N}
\end{figure}

The ratio $n_1/N$ (Table \ref{tab_exep}) shows how many threads
are created as a fraction of all posted messages. A small value
indicates that internet users are focused on the existing threads
and they are prone to continuing the previous discussions. Large
values show that there is almost no discussion, users place an
offer or question and expect only answer to them. A related
parameter that describes a discussion is the average distance from
the root node $<r>$ (Table \ref{tab_exep}). For small value of
$<r>$ the discussion is not engaging and users probably just
exchange information. For large $<r>$ vigorous discussions are
taking place. The ratio $n_1/N$ describes the behaviour of the
internet users and the average distance $<r>$ describes the
topological consequences of this behaviour. There is a functional
dependance between them and Fig. \ref{avR_n1N} demonstrates this.
The values of $n_1/N$ and $<r>$ show the kind of discussion we
examined, technical, where people are interested only in
exchanging goods, information and look for help or theoretical,
where people introduce ideas, share opinions and argue with
others. A good example are two news groups Games and Games.CS. The
Games news group is a general discussion about games, where $<r>$
is rather small. The news group Games.CS is dedicated to only one
game fans, \emph{Counter Strike} and its value of $<r>$ is much
higher than for Games news group, which suggests that the fans are
more strongly engaging within the discussion.

\subsection{The supremacy function $s(k)$}

The supremacy $s_i$ of node $i$ is defined as the total number of
all nodes that are not older than $i$ and can be linked to it by a
directed path (including the node $i$). For tree-like networks
this means that the supremacy $s_i$ of node $i$ is the total
number of nodes that are \emph{under} the node $i$, including node
$i$. In other words the supremacy $s_i$ is the total number of
nodes in the sub-tree started by node $i$. The supremacy function
$s(k)$ is the average supremacy of all nodes of degree $k$. In
\cite{Holyst} it was shown that for the Barab\'{a}si - Albert
model \cite{Barabasi_base},

\begin{equation}
  s(k)={m\over {m+1}}\left({k\over m}\right)^{m+1}+{1\over {m+1}}
\label{equ_sk1}
\end{equation}

where $m$ is a number of links created by an incoming node, and
for trees, when $m=1$
\begin{equation}
  s(k)={1\over 2}k^2+{1\over 2}.
\label{equ_sk2}
\end{equation}

\begin{figure}
\centering \subfigure {\psfig{figure=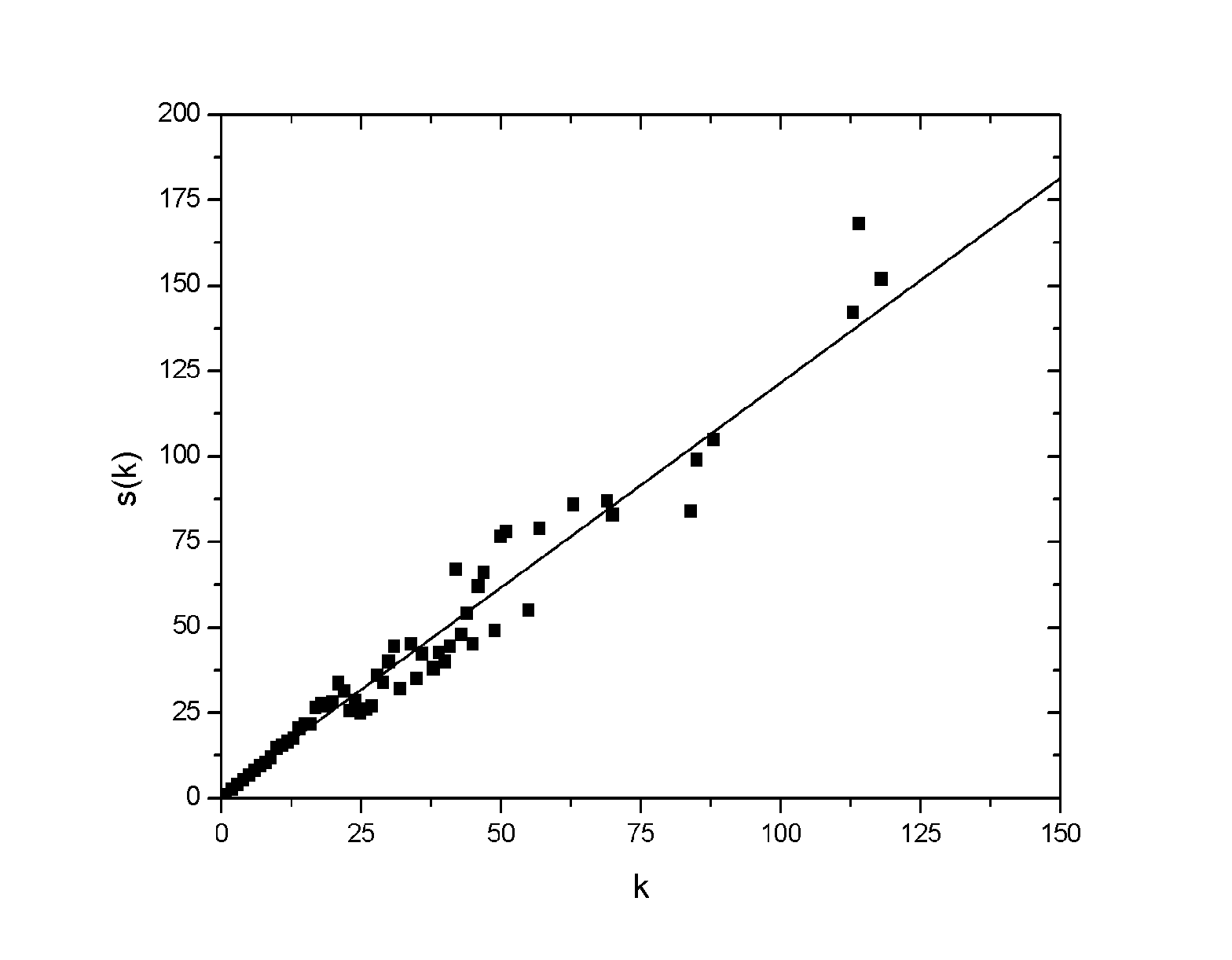,height=4.0cm}}
\subfigure {\psfig{figure=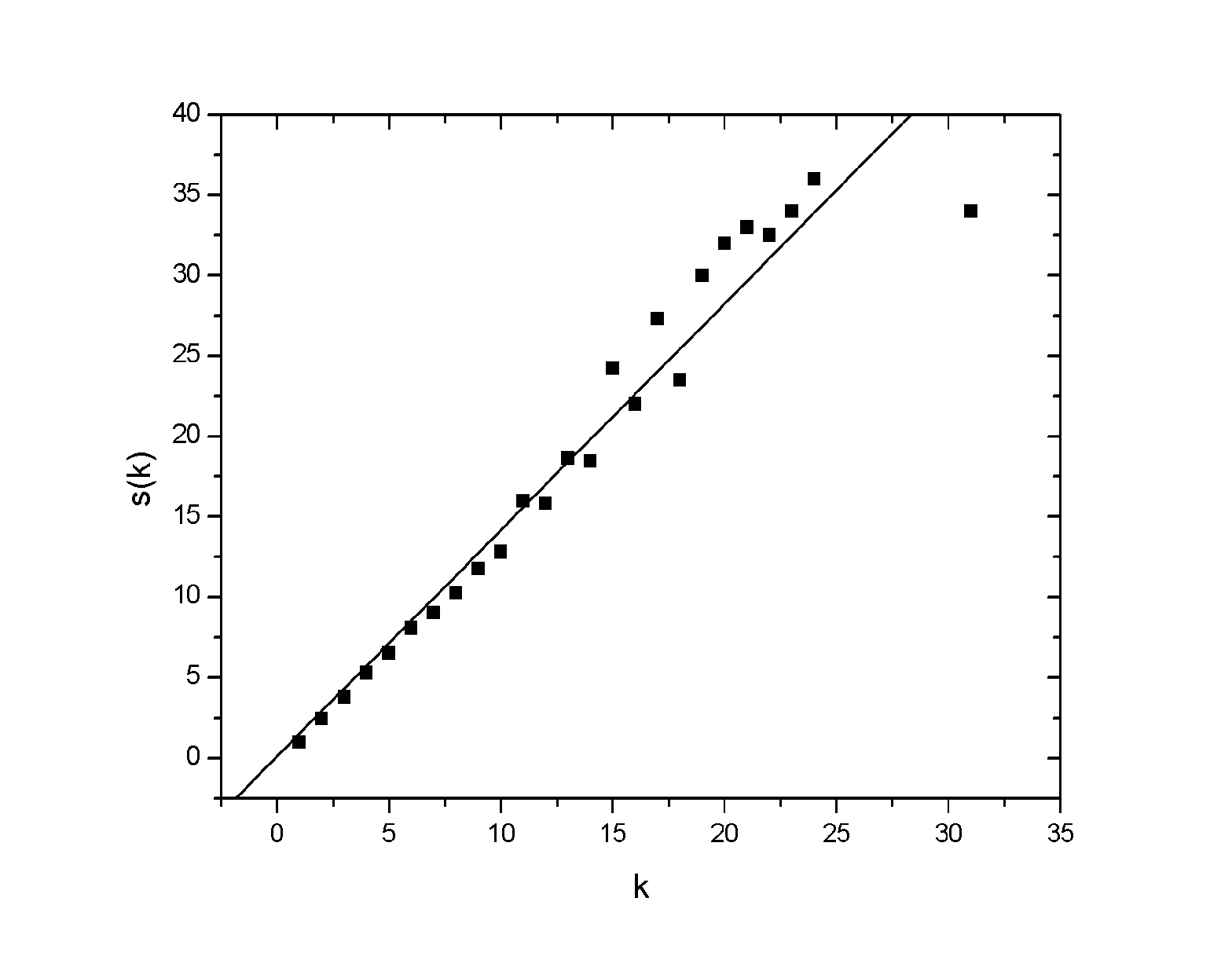,height=4.0cm}}
\caption{Average supremacy $s(k)$ against degree $k$ for (a) the
forum Poland in the EU and (b) the news group Humor. (a) and (b)
both follow linear functions with slopes $1.19$ and $1.41$
respectively. } \label{SK_dist}
\end{figure}

For each network we measured the average value $s(k)$ for a
particular degree $k$. Fig. \ref{SK_dist} shows that for the
internet discussions relation $s(k)$ is not $s\sim k^2$, but
relation is linear $s\sim k$. The result $s\sim k^2$, obtained for
Barab\'{a}si - Albert model, which does not include aging of
nodes. This suggests that the linear dependence between supremacy
$s$ and degree $k$ could be triggered by the aging of nodes.

\section{Summary}

Internet discussions are tree-like networks, whose degree
distributions are described by a power law function. The networks
are growing in time and because the posted messages become out of
date naturally, the nodes are aging. For news groups the
distribution of the network time interval between a message and a
response has two scaling regimes. The small time interval regime
probably corresponds to responses within one session of the
discussion, from people currently on-line, what corresponds with
\emph{the burst activity} studied in \cite{Barabasi}, and the
behaviour for large time intervals is generated by messages posted
later by new users arriving on-line. For the internet forums the
time interval distribution is described by $T(\tau)\sim
[\tau+\tau_0]^{-\delta}$ and shows a smooth behaviour.

The time correlations within the activity time series show that
the activity of internet discussion users is integrated with
users` daily routines on both 12 and 24 hour scales (Fig.
\ref{corel1}). These measurements could help us to define an
optimal time of operation for people interested marketing goods or
services to internet users.

The distance distribution exhibits exponential character for most
news groups, which means that discussions are not deeply embedded
within larger tree structures. The results for internet forums on
www.onet.pl show the intervention of the software employed, which
only allows a maximum distance $r=13$ in its forums. However the
distance distributions for these groups exhibit a power law
behaviour. These results can be understood by considering the
topics of these discussions. The news groups contain mostly
contain closely defined, themed, discussions which are often very
technical and frequented by experienced users. Consequently
answers are very short and directly address the problem. Thus, the
average distance $<r>$ is small. In contrast, the internet forums
have a wide range of the users, who usually want to discuss and
argue with others. This attitude towards discussion creates large
and deep tree structures.

Internet discussions are an important source of data within social
sciences. They allow the study of the topology of social
connections and their temporal statistics
\cite{makowiec,Zhongbao,Goh,Capocci,Valverde}. Our study are
focused on the growing trees of messages, whose structure and
temporal statistics,as we have shown, are related to the subject
of the discussion and the day-to-day activities of users.
Investigating the emerge, aging and dying of topics in discussion
networks should yield data on people's interests - what people
like reading or commenting on. This should give insight into the
real dynamics of people's opinion change and exchange.

\section*{Acknowledgement}
This work was supported by two EC programmes, the Marie Curie
Early Stage Training NET-ACE (MEST-CT-2004-6724) and the NEST
project CREEN (FP6-2003-NEST-Path-012864).

\newpage

\section*{Table 1}

\begin{table}
\begin{center}
\begin{tabular}{|c|l|c|c|c|c|c|}  \hline
{\em No.}  &   {\em Topic of discussion}   &   {$N$} & {$\gamma$}
& {$r_{max}$} & $n_1/N$ & $<r>$
\\
\hline   &  \textbf{Onet Forums }   &  & & & &
\\
\hline 1 &   Poland in the EU    & 43027& 3.53 & 13 & 0.118 &
4.127
\\
\hline 2 &   Opinions of Poles    & 36479& 3.28 & 13 & 0.103 &
4.062
\\
\hline 3 &   Situation in Middle East & 47075& 3.37 & 13 & 0.048 &
5.701
\\
\hline  &   \textbf{News groups}    & & & & &
\\
\hline 1 &   Trade   & 44266& 5.23 & 24 & 0.517 & 1.905
\\
\hline 2 &   Politics    & 11706& 5.52 & 46 & 0.078 & 7.041
\\
\hline 3 &   Humor    & 52525& 3.90 & 76 & 0.204 & 3.534
\\
\hline 4 &   Off-topics    & 21940& 4.71 & 51 & 0.188 & 4.153
\\
\hline 5 &   Linux    & 11049& 4.87 & 25 & 0.208 & 3.234
\\
\hline 6 &   Pillory    & 40495& 4.70 & 62 & 0.132 & 5.299
\\
\hline 7 &   Games    & 34080& 5.37 & 30 & 0.293 & 2.811
\\
\hline 8 &   Games.CS    & 18976& 4.46 & 25 & 0.162 & 3.698
\\
\hline 9 &   Programming   & 14560& 5.50 & 25 & 0.261 & 2.948
\\
\hline 10 &   Music    & 12461& 5.49 & 20 & 0.359 & 2.481
\\
\hline 11 &   Campus.Riviera    & 15431& 5.08 & 33 & 0.326 & 2.821
\\
\hline 12 &   Campus.Ustronie    & 31170& 5.10 & 26 & 0.317 &
2.897
\\
\hline 13 &   Electronics    & 28199& 5.75 & 18 & 0.364 & 2.329
\\
\hline 14 &   Windows    & 13684& 5.84 & 32 & 0.210 & 3.575
\\
\hline 15 &   Film    & 32923& 5.16 & 20 & 0.306 & 2.783
\\
\hline
\end{tabular}

\end{center}
\caption{We measured 19 internet discussions, 4 from the internet
forum www.onet.pl and 15 news groups from the server
news.student.pw.edu.pl. The columns contain the name of the
discussion, the number of nodes $N$, the exponent $\gamma$ of the
power law degree distribution and the maximum distance $R_{max}$
from the root node. Next column contains number of threads $n_1$
over all messages $N$ and the last the average distance from the
root node $<r>$. } \label{tab_exep}
\end{table}

\end{document}